\documentclass[12pt]{article}

\usepackage{a4wide}
\usepackage{amsmath}
\usepackage{amsthm}
\usepackage{amsfonts}
\usepackage{amssymb}

\usepackage{cite}

\DeclareSymbolFont{AMSb}{U}{msb}{m}{n}
\DeclareSymbolFontAlphabet{\Bbb}{AMSb}

\newcommand{\Z}{\Bbb{Z}}

\newcommand{\ie}{i.e.\ }

\DeclareMathOperator{\tr}{tr}

\begin{document}

\thispagestyle{empty}
\vspace*{-80pt} 
{\hspace*{\fill} Preprint-KUL-TF-2000/09} 
\vspace{70pt} 
\begin{center} 
{\LARGE BEC for a Coupled Two-type \\[5pt]
 Hard Core Bosons Model }
 \\[25pt]  
 
{\large  
    J.Lauwers\footnote{Bursaal KUL FLOF-10408}\footnotetext{Email: {\tt joris.lauwers@fys.kuleuven.ac.be}},
	A.Verbeure\footnote{Email: {\tt andre.verbeure@fys.kuleuven.ac.be}}
    } \\[25pt]   
{Instituut voor Theoretische Fysica} \\  
{Katholieke Universiteit Leuven} \\  
{Celestijnenlaan 200D} \\  
{B-3001 Leuven, Belgium}\\[25pt]
{March, 2000}\\[60pt]
\end{center} 
\begin{abstract}\noindent
We study a solvable model of two types hard core Bose particles. A complete analysis
is given of its equilibrium states including the proof of existence of
Bose-Einstein condensation. The plasmon frequencies and the quantum normal
modes corresponding to these frequencies are rigorously constructed. In
particular we show a two-fold degeneracy of these frequencies. We show that
all this results from spontaneous gauge symmetry breakdown. 
\\[15pt]
\begin{center}
{\bf  PACS - Keywords}
\end{center}
03.75.F - 11.15.E - 05.40\\
Bose-Einstein condensation, spontaneous symmetry breakdown, fluctuation operators,
normal coordinates, plasmon frequencies.
\end{abstract}

\newpage

\section{Introduction} 
The study of lattice Bose gas models was proposed \cite{matsubara:1956,
matsubara:1956b} in order
to understand the effect of interaction on Bose-Einstein Condensation
(BEC). It was argumented that, if the interaction has a hard core, \ie each
site can be occupied by at most one particle, then the model can be mapped onto
a quantum spin-$\tfrac{1}{2}$ XY-model on a $\Z^d$-lattice.

The exact solution, exhibiting condensation, for the hard core Bose gas on a
complete graph has been obtained \cite{toth:1990, penrose:1991}. In spin
language this work corresponds to the mean field spin-$\tfrac{1}{2}$ XY-model.
Other exact results on the full model are an upper bound on the condensate 
\cite{toth:1991} and an existence proof of condensation for the system in 
a half filled $\Z^d$-lattice for dimensions $d \geq 3$ \cite{angelescu:1992}.
In this context we mention also the relevant result \cite{kennedy:1988} about
the occurrence of long-range order in the XY-model with spins and dimensions
greater than one.

In this paper we turn our attention to a system of two types of interacting hard
core Bose particles on a complete graph and interacting through a repulsive
interaction if the particles are of different type. (Compare this with the Hubbard
model for electron systems.)  The model is exactly soluble. We study in detail
the equilibrium states in the case of a half filled lattice and equal density
for the two particle types. If the system is weakly interacting we obtain an
explicit and complete analysis of the equilibrium states. We analyse the BEC
phenomenon and in particular we are interested in the appearance
of spontaneous symmetry breaking (SSB) linked with it. The interesting aspect 
of this model is
that the broken symmetry is two-dimensional. While, as is well known for a
single type Bose system, the broken symmetry is the one-dimensional gauge
symmetry. We compute rigorously the plasmon frequencies related to these modes,
and find the quantum normal modes of these plasmon motions. We find out that
there are two different frequencies, but that both frequencies are two-fold
degenerate. The four quantum degrees of freedom have normal modes which we build
up as the quantum fluctuation operators of the components of the infinitesimal
generators of the spontaneously broken gauge symmetries, with adjoints the
quantum fluctuation operators of corresponding order parameter operators. 
The degeneracy of the plasmon spectrum is the new
phenomenon in this model study of Bose-Einstein condensation. 
Another revealing fact is that the adjoint fluctuation operators are the total 
and the relative current fluctuations between the two types of bosons.

\section{The Model and its Equilibrium States} \label{equi}

\subsection{The Model}

The model is defined on the lattice $\Z$, by the local Hamiltonians 
$H_\Lambda$, with $\Lambda$ a finite interval in $\Z$: 
\begin{equation}\label{HL}
H_{\Lambda} = \frac{t}{|\Lambda|}\sum_{\alpha \in \{1,2\}}\ \sum_{i \neq j \in \Lambda}
\, \sigma^{+}_{i \alpha} \sigma_{j \alpha}^- -  
\mu \sum_{\alpha \in \{1,2\}}\ \sum_{i \in \Lambda}
\, \sigma^{+}_{i \alpha} \sigma_{i \alpha}^- +
U \sum_{i \in \Lambda} \, \sigma^{+}_{i 1}\sigma_{i 1}^- 
\sigma^{+}_{i 2}\sigma_{i 2}^-. 
\end{equation}
The first term is the hopping contribution, the particles jump freely from one
site to another with a hopping rate $t/|\Lambda|$, $|\Lambda|$ is the
number of points in $\Lambda$; the index $\alpha$ specifies the type of
particles; $\mu$ is the chemical potential; $U$ is taken to be positive and is
the coupling constant for the interaction between the types of particles. If two
particles of different type are on the same lattice point the potential energy
is raised with an amount $U$. Clearly the $\sigma^{\pm},\sigma^z$ are the usual
Pauli matrices and are here the creation an annihilation operators for the Bose
particles with hard core, satisfying:
\begin{align*}
  &\mathbf{[} \sigma^+_{i \alpha},\sigma^-_{j\beta}\mathbf{]} 
  = \sigma_{i\alpha}^z\delta_{ij} \delta_{\alpha \beta}; \\
  &\left(\sigma^+_{i \alpha}\right)^2 = 0; \quad \left(\sigma^+_{i
  \alpha}\right)^* = \sigma^-_{i\alpha}; \\
  &\sigma^+_{i \alpha}\sigma^-_{i \alpha} = \frac{1}{2}\big(\sigma^z_{i \alpha}
  +  \mathbf{1}\big). 
\end{align*}
We look first for the equilibrium states of this model in the thermodynamic 
limit. The Hamiltonians (\ref{HL}) are permutation invariant with respect 
to the lattice index. This means \cite{fannes:1980} that the equilibrium states
at inverse temperatures $\beta$ are convex combinations of product states on the
infinite tensor product algebra of observables $\mathcal{M} = \bigotimes_{i \in
\Z}\left( M_2 \otimes M_2 \right)_i$ of the system; $\left( M_2 \otimes 
M_2 \right)_i$ represents the tensor product of the $2 \times 2$ complex
matrices $M_2$ with itself. Each set $M_2$ is referring to a different 
type of hard core bosons, both at site $i \in \Z$. 
In particular, if one looks only at the product
states which are invariant for the interchange of the two 
types of particles, then the equilibrium states are given 
by the states $\omega_{\beta \lambda}(\,.\,)$ of the form \cite{fannes:1980}:
\begin{equation*}
  \omega_{\beta \lambda}(A_1A_2\dots A_n) = \prod_{i = 1}^{n} \tr
  \rho_{\beta\lambda} A_i; \quad \text{for any } A_i \in \left( M_2 \otimes 
  M_2 \right)_i, i = 1, \dots n
\end{equation*}
where $\rho_{\beta \lambda}$ is a density matrix in $M_2 \otimes M_2$,
satisfying the self-consistency equation:  
\begin{equation}\label{s-ce}
\rho_{\beta \lambda} = \frac{\exp [-\beta h_\lambda ]}
{\tr \exp [-\beta h_\lambda ]}, 
\end{equation}
where
\begin{equation}\label{hl1}
h_\lambda = t \lambda \left( \sigma^+_1 + \sigma^+_2\right)
 + t \bar{\lambda} \left( \sigma^-_1 + \sigma^-_2 \right)
 - \mu \left ( \sigma^+_1 \sigma^-_1 + \sigma^+_2 \sigma^-_2 \right )
 + U \sigma^+_1 \sigma^-_1  \sigma^+_2 \sigma^-_2
\end{equation}
and
\begin{equation}\label{l}
  \lambda =  \tr \rho_{\beta \lambda}\sigma^-_1 
  = \tr \rho_{\beta \lambda} \sigma^-_2.
\end{equation}
In fact equation (\ref{s-ce})  for $\rho_{\beta \lambda}$ is equivalent to the
equation (\ref{l}) for $\lambda$. Remark that (\ref{l}) also results from the
assumption of symmetry in the particle types. Looking for equilibrium states
breaking this symmetry would violate (\ref{l}).

Moreover, we will limit here our study to the case of a half filled lattice.
Mathematically this means that we require:
\begin{equation*}
  \omega_{\beta \lambda}(\sigma^z_{i\alpha}) = 0 \quad \text{for all } i \in \Z,
  \quad \alpha = 1,2.
\end{equation*}
This fixes the chemical potential $\mu = U/2$.
In order to simplify the notations, we adapt our Hamiltonian 
by adding a constant term and by scaling the constant $U$ to $U/4$. 
In this case one gets for (\ref{hl1}) a more simple expression:
\begin{equation}\label{hl}
h_\lambda = t \lambda \left( \sigma^+_1 + \sigma^+_2\right)
 + t \bar{\lambda} \left( \sigma^-_1 + \sigma^-_2 \right)
 + U \sigma^z_1 \sigma^z_2.
\end{equation}
On a finite subset $\Lambda \subset \Z$ we define Fourier-transformed
operators:
\begin{equation*}
  a_{k\alpha} = \frac{1}{\sqrt{|\Lambda|}} \sum_{j \in \Lambda}\sigma^-_j 
  \exp\left[\mathrm{i} \frac{2 \pi}{|\Lambda|}kj\right]; \quad k \in \Lambda^*\ \text{the
  dual of $\Lambda$}, 
\end{equation*}
the annihilation operator of the $\alpha$-particle with quasi-momentum $k$. 

Then, 
\begin{equation*}
  n_k  = \sum_{\alpha = 1}^2 a^*_{k\alpha} a_{k\alpha}
\end{equation*}
is the number operator of particles with quasi-momentum $k$.

Compute in a state $\omega_{\beta \lambda}$ (\ref{s-ce}):
\begin{equation}\label{rho0}
  \rho_0 = \lim_{|\Lambda| \to \infty}
  \omega_{\beta\lambda}\Big( \frac{n_0}{|\Lambda|} \Big) =
  2\omega_{\beta\lambda}(\sigma^+)\omega_{\beta\lambda}(\sigma^-) 
  = 2|\lambda|^2;
\end{equation} 
\ie the $0$-mode has a density $\rho_0$ equal to $2|\lambda|^2$.
A solution  $\lambda \ne 0$ of (\ref{l}) yields  a macroscopic 
occupation of the $0$-mode and the density of the condensate $\rho_0$ equals
$2|\lambda|^2$.
Therefore, if one finds solutions of (\ref{s-ce}) or (\ref{l}) such that 
$\lambda \ne 0$, one proved Bose-Einstein condensation.

\subsection{Equilibrium States}

Now we proceed to look for the solution of (\ref{l}) and its properties. It is
clear that based on general results \cite{fannes:1980}
we reduced our problem to a one-site problem,
namely looking for the density matrix state $\tr \rho_{\beta\lambda}\,.\,$ with 
$\rho_{\beta\lambda} \in  M_2 \otimes M_2$.

In general  $ M_2 \otimes M_2 = M_4 $ and we use the following isomorphism:
\begin{eqnarray*}
\left (
  \begin{array}{cc}
    a_{11} & a_{12} \\
    a_{21} & a_{22}
  \end{array}
\right )
\otimes
\left (
  \begin{array}{cc}
    b_{11} & b_{12} \\
    b_{21} & b_{22}
  \end{array}
\right )
= \left ( \begin{array}{cccc}
    a_{11}b_{11} & a_{12}b_{11} & a_{11}b_{12} & a_{12}b_{12} \\
    a_{21}b_{11} & a_{22}b_{11} & a_{21}b_{12} & a_{22}b_{12} \\
    a_{11}b_{21} & a_{12}b_{21} & a_{11}b_{22} & a_{12}b_{22} \\
    a_{21}b_{21} & a_{22}b_{21} & a_{21}b_{22} & a_{22}b_{22}
  \end{array} \right ).
\end{eqnarray*}

This yields the following matrix representation for $h_\lambda$:

\begin{equation}\label{hlm}
  h_\lambda = \left ( \begin{array}{cccc} 
                        U & t \lambda & t \lambda & 0 \\
                      	t \bar{\lambda} & -U & 0 & t \lambda \\
			t \bar{\lambda} & 0 & -U & t \lambda \\
			0 & t \bar{\lambda} & t \bar{\lambda} & U
		     \end{array} \right ).
\end{equation}

Then a straightforward diagonalisation of $h_\lambda$ yields the following 
spectrum:
\begin{equation}\label{spectrum}
  \epsilon_0 = -U, \quad \epsilon_1 = U, \quad \epsilon_2 = \eta,
  \quad \epsilon_3 = -\eta,
\end{equation}
where $\eta = \sqrt{U^2 + 4 |t \lambda|^2}$.

The corresponding eigenvectors are:
\begin{equation}\label{evectors}
 \begin{aligned}
 |\phi_0 \rangle &= \frac{1}{\sqrt{2}} \left ( \begin{array}{c} 0 \\ 1 \\ -1\\ 0 \end{array} 
 \right ), \\
 |\phi_2\rangle &= \frac{1}{\sqrt{\eta(\eta - U)}}\left ( \begin{array}{c} t
 \lambda \\ \frac{\eta - U}
  {2} \\\frac{\eta - U}{2} \\ t \bar{\lambda} \end{array}\right ), 
  \end{aligned}
  \qquad
  \begin{aligned}
 |\phi_1 \rangle & = \frac{1}{\sqrt{2}|t\lambda|} \left ( \begin{array}{c} t \lambda \\ 0\\ 0\\ 
  - t \bar{\lambda} \end{array}\right ), \\ 
 |\phi_3\rangle &= \frac{1}{\sqrt{\eta(\eta + U)}}\left ( \begin{array}{c} t
 \lambda \\ -\frac{\eta + U}
  {2} \\-\frac{\eta + U}{2} \\ t \bar{\lambda} \end{array}\right ).
\end{aligned}
\end{equation}

Using (\ref{spectrum}) and (\ref{evectors}) one gets for equation (\ref{l}) the
following explicit and nonlinear equation for $\lambda$:
\begin{equation} \label{gap-final}
\lambda \left[ 1 + \frac{t}{\eta} \frac{\sinh( \beta \eta )}
   { \cosh (\beta U) + \cosh (\beta \eta )} \right] = 0.
\end{equation}
Remark that $\lambda = 0$ is always a solution. It corresponds to a state
without condensation (see (\ref{rho0})).

Looking for solutions $\lambda \ne 0$ amounts to look for a solution $\eta$
satisfying:
\begin{equation}\label{etaisfeta}
  \eta  =  f_\beta(\eta);
\end{equation}
where
\begin{equation*} 
  f_\beta(\eta)  =  - t \frac{\sinh(\beta \eta)}{\cosh(\beta U) + 
  \cosh(\beta \eta)}.
\end{equation*}
Since the condensate density $\rho_0 = 2|\lambda|^2$ (cfr.\ (\ref{rho0})) 
cannot exceed the total particle density $\rho$ 
(which is equal to one in the half filled lattice), $\sqrt{U^2 +2t^2}$ 
is an upper bound on $\eta = \sqrt{U^2 + 4|t\lambda|^2}$. The minimum value 
of $\eta$ is $U$. This defines $I$, the range of accessible values for $\eta$:
\begin{equation*}
  \eta \in I = \left[ U, \sqrt{U^2 +2t^2}\right].
\end{equation*}
Remark that $f_\beta(\eta)$ is unaffected by the sign of $U$. On the other hand,
equation (\ref{etaisfeta}) has no solutions if $t \geq 0$. Therefore we take
$t<0$ and $U > 0$.
\\
\\
\textit{Discussion of the solutions of (\ref{gap-final})}
\\
\\
We are looking for solutions yielding a second order phase transition.

Remark first that for $\eta \in I$:
\begin{align}
  \label{1deriv} 
  \frac{\partial f_\beta(\eta)}{\partial \eta} & > 0 \\
  \label{2deriv}
  \frac{\partial^2 f_\beta(\eta)}{{\partial \eta}^2} & < 0 \\
  \label{bderiv}
  \frac{\partial f_\beta(\eta)}{\partial \beta} & > 0
\end{align}
\ie $\eta \to f_\beta(\eta)$ is a monotonically increasing concave
function on $I$; also $\beta \to f_\beta(\eta)$ is monotonically increasing for
$T$ decreasing.

The problem here is that, due to the presence of the potential $U$, these
properties are not valid outside the interval $I$, such that the situation is
more complicated than for the usual mean field case.

First we consider temperature $T > 0$, and we distinguish two cases:

\textit{case a):}
\begin{equation}\label{fu>u}
  f_\beta(U) > U
\end{equation}
yielding
\begin{equation}\label{vwdopbeta}
  \frac{1}{2U} \ln \left( \frac{-t + 2 U}{-t - 2 U} \right) < \beta;
\end{equation}
and implying the condition on the interparticle interaction:
\begin{equation}\label{2u<t}
  2U < -t
\end{equation}
\ie the interparticle interaction must be relatively small compared with the
hopping term (\ref{HL}); 
this condition is a meaningful situation for
this kind of systems, if the interaction becomes to strong, the typical quantum
effects like BEC disappear.

One derives also
\begin{equation} \label{fetamax<etamax}
   f_\beta\left(\sqrt{U^2 + 2t^2}\right) \leq  \sqrt{U^2 + 2t^2}.
\end{equation}
Together, condition (\ref{fu>u}) and (\ref{fetamax<etamax}) yield a unique
solution of $f_\beta(\eta) = \eta \in I$, and because of (\ref{bderiv}) 
a second order phase transition with $\lambda \ne 0$
as an order parameter. Equation (\ref{vwdopbeta}) learns that there exists a
critical $\beta_c$ determinated by:
\begin{equation}\label{betac}
  \frac{1}{k_BT_c}= \beta_c  = \frac{1}{2U} 
  \ln \left( \frac{-t + 2 U}{-t - 2 U} \right).
\end{equation}
This defines the critical temperature $T_c$.

\textit{case b):}
\begin{equation}\label{fu<u}
  f_\beta(U) < U
\end{equation}
This case is somewhat more complicated, a careful analysis learns that apart
from a second order phase transition, a first order one can occur, even if
(\ref{2u<t}) is satisfied. In order to avoid non-natural first order
transitions, one has to impose:
\begin{equation*}
  \left.\frac{ \partial f_{\beta_c}(\eta)}{\partial \eta} \right |_U \leq 1;
  \quad f_{\beta_c}(U) = U.
\end{equation*}
The analysis of these yields a stronger condition on $U$ than (\ref{2u<t}):
\begin{equation*}
 U < (-t)\kappa,
\end{equation*} 
where $\kappa$ can be computed numerically as the solution of the equation:
\begin{equation*}
 \begin{split}
  & \ln \left ( \frac{1 + 2 \kappa}{1 - 2 \kappa} \right )= \frac{4 \kappa}{1 - 2
  \kappa^2}; \\ 
  & \kappa \approx 0.461292.
 \end{split}
\end{equation*}  
Finally we consider the ground state $(T = 0)$ situation.
The self-consistency equation (\ref{gap-final}) becomes: 
\begin{equation*}
  \eta =  \lim_{\beta \to \infty} f_\beta(\eta) = f_\infty(\eta); 
\end{equation*}
with:
  \begin{equation*}
  \ f_\infty(\eta) =
  \begin{cases}
     0 & \text{if} \ \eta < U \\
     -t & \text{if} \ \eta > U.
  \end{cases}
\end{equation*}
If $U > -t$, then there are no non-trivial solutions $\lambda \ne 0$.
On the other hand, if $U < -t$, then there exists always a unique solution
$\lambda \ne 0$. It is obtained as the solution of:
\begin{equation*}
-t = \sqrt{U^2 + 4t^2|\lambda|^2}
\end{equation*}
given by: 
\begin{equation*}
|\lambda| = \frac{1}{2t}\left(t^2 - U^2 \right)^{1/2} < \frac{1}{2}.
\end{equation*}
Since $f_\beta(\eta)$ is increasing with $T$ decreasing (\ref{bderiv}),  
this also implies an expression for a
supremum on the condensate density $\rho_0$ (\ref{rho0}):
\begin{equation}
\rho_0 \leq \frac{\left(t^2 - U^2 \right)}{2t^2} < \rho.
\end{equation}
The condensate density is always strictly smaller than the total particle 
density, which is compatible with former results on this topic
 \cite{penrose:1991,toth:1991}. 
This finishes the study of the equilibrium properties of the model (\ref{HL}).
\section{Spontaneous Symmetry Breaking} \label{ssb}
The model $H_\Lambda$ (\ref{HL}) and its equilibrium states have a discrete
symmetry consisting of the interchange of the two particle types into each 
other. It is implemented by the following unitary matrix:  
$u = \sigma^+_1\sigma^-_1\sigma^+_2\sigma^-_2 +
         \sigma^-_1\sigma^+_1\sigma^-_2\sigma^+_2 + 
         \sigma^+_1\sigma^-_2 + \sigma^-_1\sigma^+_2$,
then for any local observable $A_{12}$ depending on the types $1$ and $2$, one gets:
\begin{equation*}\label{1-2}
  \gamma_{1 \leftrightarrow 2}(A_{12}) = \left( \prod_{i \in \Lambda}(u_{12})_i 
  \right ) A_{12} \left(
  \prod_{i \in \Lambda}(u_{12})_i \right) = A_{21}.
\end{equation*}

The eigenvectors (\ref{evectors}) of $h_\lambda$ satisfy
\begin{equation*}
u_{12}| \phi_0 \rangle = - | \phi_0 \rangle; \quad
u_{12}| \phi_1 \rangle =  \ | \phi_1 \rangle; \quad
u_{12}| \phi_2 \rangle =  \ | \phi_2 \rangle; \quad
u_{12}| \phi_3 \rangle =  \ | \phi_3 \rangle. 
\end{equation*}
A second discrete symmetry is the particle-hole ($p \!\leftrightarrow \!h$) symmetry. This symmetry
interchanges the creation and annihilation operators ($\sigma^+_i$ and
$\sigma^-_i$) and is implemented by the
unitaries $\sigma^x_{12} = \sigma^x_1 \sigma^x_2$ ($\sigma^x$ is the
x-component Pauli matrix). The invariance of $H_\Lambda$ (\ref{HL}) is expressed
by:
\begin{equation*}\label{p-h}
 \begin{split}
  \gamma_{p \leftrightarrow h}(H_\Lambda) &= \left( \prod_{i \in \Lambda}
  (\sigma^x_{12})_i \right)H_\Lambda \left( \prod_{i \in \Lambda}
  (\sigma^x_{12})_i \right) \\
  & = H_\Lambda.
 \end{split}
\end{equation*}
The eigenfunctions (\ref{evectors}) of $h_\lambda$ satisfy
\begin{equation*}
\sigma^x_{12} | \phi_0 \rangle  = - | \phi_0 \rangle;
\quad
\sigma^x_{12} | \phi_1 \rangle  = - | \phi_1 \rangle;
\quad
\sigma^x_{12} | \phi_2 \rangle  = \ | \phi_2 \rangle;
\quad
\sigma^x_{12} | \phi_3 \rangle  = \ | \phi_3 \rangle.
\end{equation*}

The Hamiltonian $H_\Lambda$ (\ref{HL}) has also a two-dimensional continuous 
gauge symmetry group given by the actions:
\begin{equation}\label{gauge}
  \gamma_{\phi_1\phi_2}(\sigma^+_k) = \mathrm{e}^{\mathrm{i}\phi_k}\sigma^+_k,
  \quad \phi_k \in [0,2\pi], \quad k = 1,2.
\end{equation}
For a finite volume $\Lambda$, it is implemented by the unitaries:
\begin{equation*}
  U_{\Lambda(\phi_1,\phi_2)} = \exp \frac{\mathrm{i}}{2}
  \Big(\phi_1 Q_{1,\Lambda}+ \phi_2 Q_{2,\Lambda}\Big);
\end{equation*}
where
\begin{equation*}
  Q_{k,\Lambda} = \sum_{j \in \Lambda}(\sigma^z_k)_j; \quad k = 1,2.
\end{equation*}
Clearly, $ \gamma_{\phi_1\phi_2}(H_\Lambda) = U_{\Lambda(\phi_1,\phi_2)}
H_\Lambda U^*_{\Lambda(\phi_1,\phi_2)}=  H_\Lambda$. Hence, per lattice site one
has a commutative two-dimensional Lie-algebra of this symmetry group,
generated by $\sigma^z_1$ and $\sigma^z_2$. However, the equilibrium states 
$\omega_{\beta \lambda}$ with $\lambda \ne 0$ (see section \ref{equi}) break
this symmetry down.
Indeed we have here for $k = 1,2$:
\begin{equation*}
 \omega_{\beta\lambda}\left(\gamma_{\phi_1 \phi_2}\left(\sigma_k^+\right)\right) = 
 \mathrm{e}^{-\mathrm{i}\phi_k}
 \omega_{\beta\lambda} (\sigma_k^+)\neq \omega_{\beta\lambda} (\sigma_k^+)
\end{equation*}
In contrast to all other models for which we have been studying the Goldstone
phenomenon \cite{michoel:1999b,michoel:2001}, here we have the spontaneous
symmetry breakdown of a two-dimensional group.
In the following, we construct again the canonical normal modes of the Goldstone
bosons together with their corresponding plasmon frequencies.
We take a symmetry breaking state $\omega_{\beta\lambda}$ with $\lambda 
\ne 0$  and we consider fluctuation mode operators. Before constructing the
relevant canonical normal coordinates of the Goldstone particles in the next
section, we introduce here the notion of \textit{fluctuation operators}. 

For the mathematical details we refer to \cite{goderis:1989,goderis:1990}.
Denote by $M_4^{sa} = (M_2 \otimes M_2)^{sa}$ the vectorspace of the Hermitian
$4 \times 4$ matrices, \ie these are one-site observables.  
\newline
For any $A \in
M_4^{sa}$, its fluctuation operator $F_\Lambda(A)$ for the volume $\Lambda$
in the state $\omega_{\beta\lambda}$ is as usual defined by:
\begin{equation*}
  F_{\Lambda,\lambda}(A) = \frac{1}{\sqrt{|\Lambda|}} \sum_{x \in \Lambda}
  \left(A_x - \omega_{\beta\lambda}(A)\right),
\end{equation*} 
where $A_x$ is a copy of $A$ on site $x \in \Z$. In \cite{goderis:1989} it is
proved that the limit:
\begin{equation*}
  \lim_{\Lambda \to \Z} F_{\Lambda,\lambda}(A) \equiv F_\lambda(A)
\end{equation*}
exists in the sense of a non-commutative central limit as an operator on a
Hilbert space $\mathcal{H}_\lambda$, generated by a normalised vector
$\tilde{\Omega}_\lambda$, and vectors $F_\lambda(A_1)\ldots
F_\lambda(A_n)\tilde{\Omega}_\lambda$, with $A_i \in  M_4^{sa}$, $n$ 
is arbitrary and a scalar product: 

\hspace*{10mm} for $A_i, B_j \in M_4^{sa}$,
\begin{equation}\label{sp}
  \Big( F_\lambda(A_1) \ldots F_\lambda(A_n) \tilde{\Omega}_\lambda,
  F_\lambda(B_1)\ldots F_\lambda(B_m)\tilde{\Omega}_\lambda\Big)
   = \delta_{nm} \mathrm{Perm.}\! \left( (\tilde{\Omega}_\lambda,F_\lambda(A_i)
   F_\lambda(B_j)\tilde{\Omega}_\lambda) \right)_{i,j = 1,\ldots n}
\end{equation}

where the two-point function is given by: 
\begin{equation}\label{2pt}
  \left(\tilde{\Omega}_\lambda,F_\lambda(A_i)
  F_\lambda(B_j)\tilde{\Omega}_\lambda \right) 
  = \omega_{\beta\lambda}(AB) -\omega_{\beta\lambda}(A)\,\omega_{\beta\lambda}(B).
\end{equation}
\ie by the truncated two-point functions of the given state 
$\omega_{\beta\lambda}$.
Furthermore, as already could be understood from (\ref{sp}), the fluctuation
operators $F_\lambda(A),F_\lambda(B),\ldots$ satisfy the canonical commutation
relations:
\begin{equation}\label{ccr}
  \mathrm{[} F_\lambda(A), F_\lambda(B)\mathrm{]}  = \omega_{\beta\lambda}(
  \mathrm{[} A,B \mathrm{]}) \mathbf{1}
\end{equation}
\ie the fluctuation operators are quantum canonical variables
(compare $\mathrm{[} q,p \mathrm{]}= \mathrm{i}\hbar$) with quantisation
parameter $ \omega_{\beta\lambda}(\mathrm{[} A,B \mathrm{]})$  (instead of 
$\mathrm{i}\hbar$) and the state $\left(\tilde{\Omega}_\lambda,. \,
\tilde{\Omega}_\lambda \right)$ is a generalised free or quasi-free state of 
a boson field algebra of observables. In section \ref{goldstone}, we look for the relevant
operators $A,B,\ldots$ in order to obtain the normal canonical variables for the
Goldstone particles, \ie the particles generated by the spontaneous
symmetry breaking. We derive the dynamical equations for these modes and we solve them.

\section{Spontaneous Symmetry Breaking and Plasma Oscillations}
\label{goldstone}

We will consider fluctuations of the generators
\begin{equation*}
  Q_{\pm} = \sigma^z_1 \pm \sigma^z_2
\end{equation*}
of the broken symmetry. 

Let $P_i, \ (i=0,1,2,3)$ be the spectral projections on the eigenfunctions
$|\phi_i \rangle$ corresponding to the eigenvalues $\epsilon_i$ (\ref{spectrum})
of $h_\lambda$ (\ref{hlm}), \ie $h_\lambda = \sum_{i=0}^3 \epsilon_i P_i$.
Consider for any local observable $A \in M_4^{sa}$ the Hermitian operators:
\begin{equation}\label{E}
   E_{ij}(A)  = P_i A P_j + P_j A P_i; \qquad 
  i < j; \quad i,j=0,1,2,3;
\end{equation}   
  and the adjoints:
\begin{equation}\label{JE}  
  JE_{ij}(A)  = \mathrm{i}\left(P_i A P_j -  P_j A P_i \right). 
\end{equation}

An immediate calculation leads to:
\begin{equation}\label{chl}
 \begin{split}
  &\mathrm{[}h_\lambda , E_{kl}(A) \mathrm{]}  
  =  - \mathrm{i}(\epsilon_k -\epsilon_l) JE_{kl}(A); \\  
  &\mathrm{[}h_\lambda , JE_{kl}(A) \mathrm{]}
  =  \mathrm{i}(\epsilon_k -\epsilon_l)E_{kl}(A).
 \end{split}
\end{equation}
Now we take $A = Q_+$, respectively $A = Q_-$. Due to the 
particle-hole and type-exchange symmetries (section \ref{ssb}),
the only non-zero operators of the
type (\ref{E}-\ref{JE}) are:
\begin{equation} \label{nonzeroEJE}
  E_{0i}(Q_{-}),\ JE_{0i}(Q_{-}),\ E_{1i}(Q_{+}),\ JE_{1i}(Q_{+})\quad
  i=2,3.
\end{equation}
It is interesting to write out these operators in terms of the Pauli matrices 
$\sigma^{\pm}_i,\sigma^z_i,\ i=1,2$.
As an example we give the expression for the first pair of these operators
$ \left(E_{02}(Q_-),JE_{02}(Q_-)\right)$, the others are
left as an exercise for the reader:
\begin{align*}
E_{02}(Q_-) & = \frac{\xi_-}{2\eta}Q_- + \frac{t\lambda}{\eta}(\sigma^z_1
  \sigma^+_2 - \sigma^z_2 \sigma^+_1) + \frac{t\bar{\lambda}}{\eta}(\sigma^z_1
  \sigma^-_2 - \sigma^z_2 \sigma^-_1);\\
  JE_{02}(Q_-) & = \mathrm{i}\frac{\xi_-}{2\eta}( \sigma^+_1
  \sigma^-_2 - \sigma^+_2 \sigma^-_1) + \mathrm{i}\frac{t\lambda}{\eta}
  (\sigma^+_1 - \sigma^+_2) - \mathrm{i}\frac{t\bar{\lambda}}{\eta}
  (\sigma^-_1 - \sigma^-_2).
\end{align*}
$E_{02}(Q_-)$ is a component of the infinitesimal generator
$Q_-$ and $JE_{02}(Q_-)$ is the adjoint order parameter operator to $E_{02}(Q_-)$.

It is clear from (\ref{spectrum}) that the dynamical equations (\ref{chl}) for
the operators (\ref{nonzeroEJE}) contain only the relative energies
$\epsilon_k - \epsilon_l$ given by:
\begin{equation*}
  \xi_\pm = \eta \pm U.
\end{equation*}
In particular, the energy gap $\xi_+$ is entering in the
equations:
\begin{gather}
 \label{chl02}
  \mathrm{[}h_\lambda , E_{02}(Q_-) \mathrm{]}
  = \mathrm{i}\xi_+ JE_{02}(Q_-),  \quad
  \mathrm{[}h_\lambda , JE_{02}(Q_-) \mathrm{]}
  = -\mathrm{i}\xi_+ E_{02}(Q_-),
   \\
 \label{chl13}
  \mathrm{[}h_\lambda , E_{13}(Q_+) \mathrm{]}
  = -\mathrm{i}\xi_+ JE_{13}(Q_+), \quad
  \mathrm{[}h_\lambda , JE_{13}(Q_+) \mathrm{]}
  = \mathrm{i}\xi_+ E_{13}(Q_+);
\end{gather}
and the energy gap $\xi_-$ in:
\begin{gather}
 \label{chl03}
  \mathrm{[}h_\lambda , E_{03}(Q_-) \mathrm{]}
  = -\mathrm{i}\xi_- JE_{03}(Q_-), \quad
  \mathrm{[}h_\lambda , JE_{03}(Q_-) \mathrm{]}
  = \mathrm{i}\xi_- E_{03}(Q_-),
 \\
 \label{chl12}
  \mathrm{[}h_\lambda , E_{12}(Q_+) \mathrm{]}
  = \mathrm{i}\xi_- JE_{12}(Q_+), \quad
  \mathrm{[}h_\lambda , JE_{12}(Q_+) \mathrm{]}
  = -\mathrm{i}\xi_- E_{12}(Q_+).
\end{gather}
All these results are on the level of the microscopic observables. In particular
one discovers in (\ref{chl02}-\ref{chl12}) that the dynamics leaves invariant the
pairs $ \left(E_{02}(Q_-),JE_{02}(Q_-)\right)$,
$\left(E_{13}(Q_+),JE_{13}(Q_+)\right)$,
$ \left(E_{03}(Q_-),JE_{03}(Q_-)\right)$ 
and $\left(E_{12}(Q_+),JE_{12}(Q_+)\right)$.
On the other hand the different pairs do not form canonical pairs of variables.
Therefore, we have to consider the macroscopic variables, namely  the
fluctuations of these quantities. We know from section \ref{ssb} that the
fluctuation operators are canonical variables, satisfying the canonical
commutation relations.

We treat explicitly the first pair appearing in the equation (\ref{chl02}), and
give the results for the other pairs without technical details.
\\
\\
\textit{Plasmon frequency $\xi_+$}  
\\
\\
With the notations of section \ref{ssb}, denote the fluctuations:
\begin{xalignat}{2} \label{xp02}
  X_{02} &\equiv -\frac{1}{n_{02}} F_\lambda\left(E_{02}(Q_-)\right); &
  P_{02} &\equiv \,\frac{1}{n_{02}} F_\lambda\left(JE_{02}(Q_-)\right),
\end{xalignat}
where
\begin{equation*}
  n_{02} = \left( \frac{\mathrm{e}^{\beta U} +\mathrm{e}^{-\beta
  \eta}}{Z}\right)^{1/2}; \quad Z = 2\cosh(\beta U) + 2\cosh(\beta \eta).
\end{equation*}

These satisfy (following (\ref{ccr})) the canonical commutation relation:
\begin{equation} \label{cxp02}
  \mathrm{[} X_{02},P_{02} \mathrm{]} = - \frac{1}{n^2_{02}}
  \omega_{\beta \lambda}\left( \mathrm{[}E_{02}(Q_-),JE_{02}(Q_-) 
  \mathrm{]}\right)\mathbf{1}
  = \mathrm{i} \hbar_+
\end{equation}
with quantisation parameter:
\begin{equation}\label{hbar+}
  \hbar_+ = \frac{4 \xi_-}{\eta} \tanh \left(\beta \xi_+/2\right). 
\end{equation} 
Following (\ref{2pt}), the calculations for the variances are straightforward:
\begin{equation}\label{v02}
  \left(\tilde{\Omega}_\lambda,X_{02}^2\tilde{\Omega}_\lambda\right)=
  \left(\tilde{\Omega}_\lambda,P_{02}^2\tilde{\Omega}_\lambda\right)=
  2\frac{\xi_-}{\eta}.
\end{equation}
Now compute the dynamics of these fluctuation operators induced by the
microdynamics determined by the Hamiltonian $h_\lambda$ (\ref{hl}). In
differential form the dynamics follows immediately from the equation
(\ref{chl02}):
\begin{gather*}
   \frac{d}{i dt}\left( F_\lambda\left(E_{02}(Q_-)\right)\right)
   = F_\lambda\left(\mathrm{[} 
   h_\lambda,E_{02}(Q_-)\mathrm{]}\right)
   =\mathrm{i} \xi_+ F_\lambda\left(JE_{02}(Q_-)\right);
   \\
   \frac{d}{i dt}\left( F_\lambda\left(JE_{02}(Q_-)\right)\right) 
   = F_\lambda\left(\mathrm{[}
   h_\lambda,JE_{02}(Q_-)\mathrm{]}\right) 
   = -\mathrm{i} \xi_+ F_\lambda\left(E_{02}(Q_-)\right).
\end{gather*} 
Using the notations (\ref{xp02}) and in integrated form one gets:
\begin{equation}\label{t02}
 \begin{split}
   X_{02}(t) = &
  \, X_{02}(0) \cos(\xi_+t) + P_{02}(0)\sin(\xi_+t); \\
   P_{02}(t) = &
  -X_{02}(0)\sin(\xi_+t)+ P_{02}(0)\cos(\xi_+t). 
 \end{split}
\end{equation}
\ie the pair of macroscopic variables $\left( X_{02},P_{02}\right)$ is a
canonical pair (\ref{cxp02}), evolving in time, independently from all other
degrees of freedom of the system, as a pair of quantum harmonic oscillator
variables oscillating with a frequency $\xi_+$. Remark that if the temperature
$T$ tends to the critical temperature $T_c$, \ie when $|\lambda| \to 0$, then
$\xi_- \to 0$. It follows from the equations (\ref{v02}) that the pair disappears
completely \ie $X_{02}= P_{02} = 0$ for $T \geq T_c$.

This finishes the study of the first pair of fluctuation variables.

In a completely analogous way one defines the second pair of canonical
fluctuation operators:
\begin{xalignat}{2}\label{xp13}
  X_{13} &= \frac{1}{n_{13}} F_\lambda\left(E_{13}(Q_+)\right); &
  P_{13} &= \frac{1}{n_{13}} F_\lambda\left(JE_{13}(Q_+)\right);
\end{xalignat}
with
\begin{equation*}
  n_{13} = \left( \frac{\mathrm{e}^{-\beta U} +\mathrm{e}^{\beta
  \eta}}{Z}\right)^{1/2}.
\end{equation*}
One gets, comparable to (\ref{cxp02}), (\ref{v02}) and (\ref{t02}):
\begin{equation} \label{cxp13}
  \mathrm{[} X_{13},P_{13} \mathrm{]}  
  = \mathrm{i} \hbar_+;
\end{equation}
\begin{equation}\label{v13}
  \left(\tilde{\Omega}_\lambda,X_{13}^2\tilde{\Omega}_\lambda\right)=
  \left(\tilde{\Omega}_\lambda,P_{13}^2\tilde{\Omega}_\lambda\right)=
  2\frac{\xi_-}{\eta};
\end{equation}
\begin{equation}\label{t13}
 \begin{split}
   X_{13}(t) = &
  \, X_{13}(0) \cos(\xi_+t) + P_{13}(0)\sin(\xi_+t); \\
   P_{13}(t) = &
  -X_{13}(0)\sin(\xi_+t)+ P_{13}(0)\cos(\xi_+t). 
\end{split}
\end{equation}
\\
\\
\textit{Plasmon frequency $\xi_-$}
\\
\\
Following (\ref{chl03}), consider the third pair of canonical fluctuation
operators
\begin{xalignat}{2} \label{xp03}
  X_{03} &\equiv \frac{1}{n_{03}} F_\lambda\left(E_{03}(Q_-)\right); &
  P_{03} &\equiv \frac{1}{n_{03}} F_\lambda\left(JE_{03}(Q_-)\right),
\end{xalignat}
where
\begin{equation*}
  n_{03} = \left( \frac{\mathrm{e}^{\beta U} +\mathrm{e}^{\beta
  \eta}}{Z}\right)^{1/2}.
\end{equation*}
One gets
\begin{equation} \label{cxp03}
  \mathrm{[} X_{03},P_{03} \mathrm{]}  
  = \mathrm{i} \hbar_-
\end{equation}
with a new quantisation parameter,
\begin{equation}\label{hbar-}
  \hbar_- = \frac{4 \xi_+}{\eta} \tanh \left(\beta \xi_-/2\right); 
\end{equation} 
variances,
\begin{equation}\label{v03}
  \left(\tilde{\Omega}_\lambda,X_{03}^2\tilde{\Omega}_\lambda\right)=
  \left(\tilde{\Omega}_\lambda,P_{03}^2\tilde{\Omega}_\lambda\right)=
  2\frac{\xi_+}{\eta};
\end{equation}
and dynamics
\begin{equation}\label{t03}
 \begin{split}
   X_{03}(t) = &
  \, X_{03}(0) \cos(\xi_-t) + P_{03}(0)\sin(\xi_-t); \\
   P_{03}(t) = &
  -X_{03}(0)\sin(\xi_-t)+ P_{03}(0)\cos(\xi_-t). 
\end{split}
\end{equation}
Finally, following (\ref{chl12}), consider the fourth pair of canonical
fluctuation operators:
\begin{xalignat}{2} \label{xp12}
  X_{12} &\equiv -\frac{1}{n_{12}} F_\lambda\left(E_{12}(Q_+)\right); &
  P_{12} &\equiv \,\frac{1}{n_{12}} F_\lambda\left(JE_{12}(Q_+)\right),
\end{xalignat}
where
\begin{equation*}
  n_{12} = \left( \frac{\mathrm{e}^{-\beta U} +\mathrm{e}^{-\beta
  \eta}}{Z}\right)^{1/2}.
\end{equation*}
One computes the commutator:
\begin{equation} \label{cxp12}
  \mathrm{[} X_{12},P_{12} \mathrm{]}  
  = \mathrm{i} \hbar_-;
\end{equation}
the variances,
\begin{equation}\label{v12}
  \left(\tilde{\Omega}_\lambda,X_{12}^2\tilde{\Omega}_\lambda\right)=
  \left(\tilde{\Omega}_\lambda,P_{12}^2\tilde{\Omega}_\lambda\right)=
  2\frac{\xi_+}{\eta};
\end{equation}
and the dynamics
\begin{equation}\label{t12}
 \begin{split}
   X_{12}(t) = &
  \, X_{12}(0) \cos(\xi_-t) + P_{12}(0)\sin(\xi_-t); \\
   P_{12}(t) = &
  -X_{12}(0)\sin(\xi_-t)+ P_{12}(0)\cos(\xi_-t). 
\end{split}
\end{equation}

This completes the construction of the plasmon spectrum together with all its
quantum normal modes. The next question is: are these four pairs
(\ref{xp02},\ref{xp13},\ref{xp03},\ref{xp12}) all different ?\  The question is
relevant, because the mode variables are all fluctuations and coarse graining
might make some of them equivalent, \ie although $A,B \in M_4^{sa}$ and $A \ne
B$, $F_\lambda(A)$ might be equal to  $F_\lambda(B)$. The problem of equivalence
of fluctuation operators is considered before (see
\cite{goderis:1989,goderis:1990}). It turns out that they are equivalent,
\ie $F_\lambda(A) = F_\lambda(B)$, if and only if
\begin{equation}\label{equiv}
  \left(\tilde{\Omega}_\lambda,F_\lambda(A-B)^2\tilde{\Omega}_\lambda\right)=
\omega_{\beta\lambda}\left(\left(A-B-\omega_{\beta\lambda}\left(A-B\right)
\right)^2\right) = 0.
\end{equation}
If we take for $A,B$ the operators $E_{ij}(Q_\pm)$ and $JE_{ij}(Q_\pm)$ with $i
<j$, then clearly
\begin{equation*}
  \omega_{\beta\lambda}(E_{ij}(Q_\pm)) =\omega_{\beta\lambda}(JE_{ij}(Q_\pm)) =0
\end{equation*}
and we have to check
\begin{equation}\label{equiv2}
 \omega_{\beta\lambda}\left(\left(A-B\right)^2\right) =
 \omega_{\beta\lambda}\left(A^2 + B^2 -AB -BA \right) = 0;
\end{equation}
for $A$ and $B$ taken out of different pairs $ \left(E_{02}(Q_-),JE_{02}(Q_-)\right)$,
$\left(E_{13}(Q_+),JE_{13}(Q_+)\right)$,
$ \left(E_{03}(Q_-),JE_{03}(Q_-)\right)$, 
$\left(E_{12}(Q_+),JE_{12}(Q_+)\right)$.
It is immediately checked that in this case always $\omega_{\beta\lambda}
\left(AB\right) = \omega_{\beta\lambda}\left(BA\right)= 0$. Hence,
(\ref{equiv2}) reduces to
\begin{equation*}
\omega_{\beta\lambda}\left(\left(A-B\right)^2\right) =
 \omega_{\beta\lambda}\left(A^2 + B^2 \right) \ne 0;
\end{equation*}
which follows from (\ref{v02}),(\ref{v13}),(\ref{v03}) and (\ref{v12}).
Hence, the four pairs of canonical fluctuation operators are all independent,
proving the two-fold degeneracy of the plasmon frequency $\xi_+$ and $\xi_-$.

Finally, remark that if the temperature $T$ tends to the critical temperature,
\ie when $\lambda \to 0$, then the pairs $\left(X_{02},P_{02}\right)$ (\ref{xp02}) and
 $\left(X_{13},P_{13}\right)$ (\ref{xp13}) disappear completely. On the other hand the
 pairs $\left(X_{03},P_{03}\right)$ (\ref{xp03}) and $\left(X_{12},P_{12}\right)$
 (\ref{xp12}) become classical variables (see (\ref{cxp03}) and (\ref{cxp12})), and
 become constants of the motions (see (\ref{t03}) and (\ref{t12})). In fact
 for $T \geq T_c$ the continuous symmetry group (\ref{gauge}) is restored. 
The fluctuation operators $X_{03}$ and $X_{12}$ are the fluctuations of the
generators of this symmetry. The fluctuation operators $P_{03}$ and $P_{12}$
are the total particle current and relative particle current fluctuations.   
For $T < T_c$ the physical interpretation of these operators remains the
same with the basic difference that they show a definite collective or
macroscopic quantum character, and a non-trivial oscillatory time behaviour.
We remind that the results derived for this model are mathematically
rigorous. In view of the recent interest in interferences between
Bose-Einstein condensates with Josephson-like coupling \cite{wright:1997,
wang:1999} an extension of our model including such a coupling raises over
as the natural thing to do next. We leave it for a future contribution.

Acknowledgement: The authors thank Tom Michoel for helpful discussions. 

\bibliographystyle{myunsrt}
\bibliography{biblio}

\end{document}